\documentstyle[12pt,aaspp4,tighten,flushrt]{article}
\def\.{\mathaccent 95}
\def\beq{\begin{equation}}
\def\ee{\end{equation}}
\def\a{\alpha}
\def\be{\beta}

\def\la{\lambda}

\def\frac#1#2{{\textstyle{{#1}\over {#2}}}}

\def\lsim{\mathrel{\rlap{\lower4pt\hbox{\hskip1pt$\sim$}}
    \raise1pt\hbox{$<$}}}
\def\gsim{\mathrel{\rlap{\lower4pt\hbox{\hskip1pt$\sim$}}
    \raise1pt\hbox{$>$}}}
\def\sqr#1#2{{\vcenter{\vbox{\hrule height.#2pt
         \hbox{\vrule width.#2pt height#1pt \kern#1pt
         \vrule width.#2pt}
         \hrule height.#2pt}}}}

\newbox\grsign \setbox\grsign=\hbox{$>$} \newdimen\grdimen \grdimen=\ht\grsign
\newbox\simlessbox \newbox\simgreatbox
\setbox\simgreatbox=\hbox{\raise.5ex\hbox{$>$}\llap
     {\lower.5ex\hbox{$\sim$}}}\ht1=\grdimen\dp1=0pt
\setbox\simlessbox=\hbox{\raise.5ex\hbox{$<$}\llap
     {\lower.5ex\hbox{$\sim$}}}\ht2=\grdimen\dp2=0pt

\def\doublespace {\smallskipamount=6pt plus2pt minus2pt
                  \medskipamount=12pt plus4pt minus4pt
                  \bigskipamount=24pt plus8pt minus8pt
                  \normalbaselineskip=24pt plus0pt minus0pt
                  \normallineskip=2pt
                  \normallineskiplimit=0pt
                  \jot=6pt
                  {\def\smallskip {\vskip\smallskipamount}}
                  {\def\medskip   {\vskip\medskipamount}}
                  {\def\bigskip   {\vskip\bigskipamount}}
                  {\setbox\strutbox=\hbox{\vrule 
                    height17.0pt depth7.0pt width 0pt}}
                  \parskip 12.0pt
                  \normalbaselines}

\font\gkvec=cmmib10                         %for boldface lowercase
\def\bomega{\hbox{{\gkvec\char33}}}                  %bold face omega

\def\lb{\langle}
\def\rb{\rangle}

\def\bw{\bar{\omega}}
\def\bv{\bar V}
\def\bB{\bar B}
\def\ts{\times}
\def\lb{\langle}
\def\rb{\rangle}
\def\curl{\nabla {\ts}}

\def\cd{\cdot}

\def\bbV{\bar {\bf V}}
\def\bfv{{\bf v}}
\def\bfvp{{\bf v}}
\def\bfV{{\bf V}}
\def\bfj{{\bf j}}

\def\bfe{{\bf e}}
\def\bfw{{\bomega}}

\def\bfb{{\bf b}}
\def\bfbp{{\bf b}}

\def\bbB{\bar{\bf B}}
\def\nb{\nabla}
\def\curl{\nb\ts}

\def\b0{b^{(0)}}
\def\v0{v^{(0)}}
\def\w0{\omega^{(0)}}
\def\bb0{\bfb^{(0)}}
\def\bv0{\bfv^{(0)}}
\def\bw0{\bfw^{(0)}}
\def\bj0{\bfj^{(0)}}

\begin{document}

\centerline{\bf 
Resolution of an ambiguity in dynamo theory and its 
consequences for back reaction studies}
\medskip
\centerline {Eric G. Blackman, Theoretical Astrophysics, Caltech 130-33, 
Pasadena CA, 91125, USA}
\centerline {and}
\centerline {George B. Field, Center for Astrophysics, 60 Garden St., 
Cambridge MA, 02139, USA}
\centerline{(submitted to ApJ)}
%\doublespace

\centerline {\bf ABSTRACT}

An unsolved problem in turbulent dynamo theory is the 
``back reaction'' problem: to what degree does the mean magnetic
field suppress the turbulent dynamo coefficients which are needed to drive
its  growth? The answer will ultimately derive from a combination of
numerical and  analytical studies. Here we show that analytic approaches to
the dynamo and back reaction problems require one to separate 
turbulent quantities into two components: those influenced by the mean
field (which are therefore anisotropic) and  those independent of the 
mean field (and are therefore isotropic), no matter how weak the mean field is.
Upon revising the standard formalism to meet this requirement, we find that:
(1) The two types of  components often appear in the same equation, so that 
standard treatments, which do not distinguish between them, are ambiguous. 
(2) The usual first-order smoothing 
approximation that is necessary to make progress in the standard treatment is
unnecessary when the distinction is made.
(3) In contrast to previous suggestions, 
the correction to the dynamo $\a$ coefficient found by 
Pouquet et al (1976) is actually independent of the mean field, and therefore
cannot be interpreted as a quenching.  

\medskip

{\bf Subject Headings}: magnetic fields; galaxies: magnetic fields;
Sun: magnetic fields; stars: magnetic fields; turbulence; accretion discs.

\vfill
\eject
\centerline {\bf 1. Introduction}

Mean-field turbulent dynamo theory is a leading 
candidate to explain the origin of 
large-scale magnetic fields in stars and galaxies (e.g. Moffatt 1978; Parker
1979; Krause \& R\"{a}dler 1980; Zeldovich et al. 1983).  This theory 
appeals to a combination of helical turbulence, differential rotation,  and
turbulent diffusion to exponentiate  an initial seed large-scale magnetic
field. Typically, the total magnetic field is broken into a large-scale or
mean component and
a small-scale or fluctuating  component, and the rate of growth of the 
mean field is sought. The
mean field    grows on a length scale much larger than the outer scale 
of the 
turbulent velocity, with a growth time much larger than the eddy turnover
time at the outer scale. Fluid helicity provides a statistical 
correlation of small scale loops
favorable to exponential field growth.  Turbulent diffusion is needed to
redistribute  the amplified mean field since Ohmic diffusion is negligible. 
By definition, the mean-field dynamo 
then belongs to the ``slow'' dynamo class (Zeldovich et al 1983).  
On the other hand, the small-scale field amplifies quickly without the aid of
diffusion, and is thus constitutes a ``fast'' dynamo (Zeldovich et al 1983).
Rapid growth of the small-scale field necessarily accompanies the 
mean-field dynamo.  Its impact upon the growth of the mean field, and the 
impact of  the mean field itself on its growth are controversial.

Dynamo theory has usually  been 
studied in the kinematic regime, wherein the turbulent and 
ordered velocity fields are assumed to be given, and the seed field is
assumed to grow and amplify without back reactions that
affect  the turbulent motions which  drive its growth.  Kinematic dynamo
theory is incomplete, as a real dynamo should
represent a solution to the nonlinear magneto-hydrodynamic (MHD)
equations, which certainly embody back reaction.  
Although numerical simulations are necessary for our
understanding of the nonlinear theory, the simplicity of
the basic mean-field dynamo formalism warrants analytic investigation of
the nonlinear MHD equations and the resulting back reaction to
understand why or why not something like the kinematic treatment  
captures the evolution of the field even in the nonlinear
regime.

Some aspects of how the growing magnetic field affects the turbulence 
have been addressed.  Kulsrud \& Anderson (1992) confirmed previous arguments
by, e.g., Cowling (1957) \& Piddington (1981) that the small-scale field 
builds up to approximate equipartition successively from the smallest to the 
largest scales in the turbulence.  However, since most of the energy in a turbulent spectrum is concentrated at the largest - or outer -  scale, the 
question of whether the outer scale reaches true  equipartition  is most
important,  since at equipartition, MHD turbulence consists largely of
Alfv\'en waves which do not contribute to dynamo action 
(Moffatt 1978); this would strongly suppress  dynamo growth.   An equally 
important, and related question is: for what value of the mean  field does 
suppression of dynamo action become important?   The simulations of e.g. 
Cattaneo \& Hughes (1996) suggest  that suppression occurs when the energy 
density in the mean magnetic field is smaller by a factor of the magnetic
Reynolds number than that implied by e.g.,  Pouquet et al. (1976).

Not only do simulations disagree on the role of the magnetic
Reynolds number in the suppression formulas,  
but so do interpretations from well-motivated analytical 
approaches such as those of  Zeldovich et al. (1983); Montgomery \& Chen 
(1984);
Gruzinov \& Diamond (1994); 
Bhattacharjee \& Yuan (1995); Kleeorin et al. (1995) and Field et al. 
(1998). The formalism of 
Bhattacharjee \& Yuan (1995), Gruzinov \& Diamond (1994) and Kleeorin et al
(1995) leads to
extreme suppression even in the weak mean-field limit.  A second class 
of formalism is exemplified by Vainshtein \& Kitchatinov (1983) and 
Montgomery \& Chen (1994) who considered an expansion in 
the mean magnetic field, and 
Blackman \& Chou (1997) who considered a similar perturbation theory in 
both the mean magnetic field and the mean velocity field.
Field et al. (1998) developed a fully nonlinear approach 
for assessing the back reaction in the case that the gradient in the  mean
field is negligible. In contrast to  Bhattacharjee \& Yuan (1995),
Gruzinov \& Diamond (1994) and Kleeorin et al.\ (1995),  but in agreement
with Kraichnan (1979),  Field et al.(1998) find that the magnetic Reynolds
number does  not strongly enter the suppression of   the alpha
dynamo coefficient. Why do different analytic approaches in such
papers concerned with back reaction produce such different results?  Here we 
show that the answer to this question depends on how one decomposes  
small-scale turbulent quantities into   components which are 
dependent, and independent of the mean fields, respectively.   
We show that these components often appear together in the same equations
and not distinguishing them leads to ambiguities, while properly distinguishing
between them allows progress even if the first-order smoothing approximation is
not satisfied.

In section 2 we review the standard  
derivation of the dynamo coefficients 
following  Moffatt (1978). 
In Section 3 we dissect the standard treatment, point out the
ambiguities, and rederive the $\a$  coefficient with a
formalism that makes the distinction between components discussed above.
In section 4 we show the effect of the revised approach on
analytic back reaction studies. We conclude in section 5.

\medskip

\centerline {\bf 2. The standard derivation}

The induction equation describing the magnetic field evolution is
\begin{equation}
\partial_{t}{\bf B} =\curl(\bfV\ts\bf B) + 
\la\nabla^{2}{\bf B},
\label{INDUCTION}
\end{equation}
where $\la$ is a constant magnetic viscosity.
Here $\bf B$ is in velocity units, obtained by dividing by $\sqrt{4\pi\rho}$;
we assume incompressibility.

The equation for the mean field
derived by averaging  (\ref{INDUCTION})   is
\begin{equation}
\partial_{t}\bbB =
\curl\lb\bfv \ts \bfb\rb
-\bbV\cd\nabla\bbB
+\bbB\cdot\nabla\bbV
+\la\nabla^{2}\bbB,
\label{BBAR}
\end{equation}
where an overbar indicates the mean field and the small-scale quantities
are in lower case.  The term $\bbB\cdot \nabla\bbV$
describes the effects of differential rotation, and will not be discussed
further here, while the term $\bbV\cd\nb\bbB$ can be eliminated by changing 
the frame of reference to one moving with $\bbV$; both terms will be ignored
in what follows. The task for the dynamo theorist is to find the dependence of
the  turbulent EMF
$\lb\bfv\ts\bfb\rb$ on $\bbB$ so that
(\ref{BBAR}) can be solved. The kinematic 
theory assumes that the statistical properties of the small-scale
velocity field are prescribed independently of the mean
magnetic field; in that case it remains only to obtain the statistical 
behavior of the
fluctuating magnetic field. 
Subtracting  (\ref{BBAR}) from (\ref{INDUCTION})
then yields 
\begin{equation}
\begin{array}{r}
\partial_{t} \bfb = 
%\bfbp\cdot\nb\bbv
%-\bbv\cdot\nabla\bfbp
\bbB\cdot\nb\bfv-\bfv\cdot\nb\bbB
+ \bfb\cdot\nabla\bfv -\bfv\cdot\nabla\bfb
-\curl\lb\bfv\ts\bfb\rb 
+ \la\nabla^{2}\bfb.
\end{array}
\label{BPRIME}
\end{equation}

At this stage, it is customary to make the first order smoothing 
approximation (FOSA), namely, that the terms second order in the fluctuating 
quantities can be neglected.  The result is then 
\begin{equation}
\begin{array}{r}\partial_{t} \bfb = \bbB\cdot\nb\bfv-\bfv\cdot\nb\bbB
+ \la\nabla^{2}\bfb,
\end{array}
\label{BPRIME1a}
\end{equation}
and for small $\la$ the last term can be ignored in the kinematic
theory.  We use this equation to describe the statistics of $\bf b$ in terms
of those of $\bf  v$. In doing so we   employ the  Reynolds relations
(\cite{RAD80}), i.e. that derivatives with respect to $\bf x$ or $t$
obey $\partial_{t,{\bf x}}\lb { X_i X_j} \rb =
\lb\partial_{t,{\bf x}}({X_i X_j})\rb$ and $\lb
\bar{X}_{i}x_{j}\rb = 0$
where $X_i={\bar X}_i+x_i$ are
components of vector functions of $\bf x$ and $t$.  
For statistical ensemble means, these hold when  
correlation times  are small compared to the  
times over which mean quantities vary.  For spatial means,
defined by $\lb X_{i}({\bf x},t)\rb = V^{-1} \int X_{i} ({\bf x}+{\bf s},t) {\rm d}{\bf s}$,
the relations hold when the average is over a large enough
$V$ that $L \ll V^{1 \over 3} \ll D$, where $D$ is the size of the system
and $L$ is the outer scale of the turbulence.

Using the Reynolds rules, isotropy of the turbulent
velocities, and the assumption that
$\lb\bfv(t)\ts \bfb(0)\rb=0$, we obtain the classical result (Moffat 1978):
\begin{equation}
\lb\bfv\ts\bfb\rb=\lb\bfv(t)\ts\int^t_0\partial_{t'} 
\bfb dt' \rb
%+\lb\int_0^t d_{t'} 
%\bfvp dt' \ts \bfbp(t)\rb,
=\alpha \bbB-\beta \curl\bbB,
\label{EMF1}
\end{equation}
where $\a=-{1 \over 3} t_c\lb\bfv\cdot\curl\bfv\rb$
and $\be={1 \over 3}t_c\lb\bfv\cdot\bfv\rb$ are the helicity and diffusion
dynamo coefficients respectively, and $t_c$ is the correlation
time of the turbulent quantities. Eq. (\ref{EMF1}) 
 is put into (\ref{BBAR}) to solve for the evolution of the mean
field, often assuming $\nabla\a=\nabla\beta=0$. 
\medskip

\centerline {\bf 3. Including the back reaction and improving the standard 
formalism}

There are three fundamental limitations with the derivation of section 2:
(a) The kinematic approximation, according to which the velocity field
is prescribed, ignoring the back reaction of both the
small- and large-scale  magnetic field.
(b) The FOSA, according to which 1st and 2nd terms in (3) 
dominate the 3rd and 4th terms, even though it is known (Kulsrud and
Anderson 1992) that the small-scale field $b$ builds up much faster
than the large-scale field $\bar B$. (c) Finally, even though the same symbols
$\bfv$ and
$\bfb$ are employed throughout the derivation, they mean different things
in different places.  To see why, note that the scalar forms of $\a$ and
$\beta$ follow  from isotropy of the fluctuating components $\bfb$ and
$\bfv$. However, if both are strictly
isotropic, $\langle {\bf v}\times {\bf b}\rangle$ would vanish 
because it is the average of an isotropic vector.  Moreover, growth of the
mean field in (\ref{BBAR}) requires some violation of homogeneity, because
growth requires a non-zero
$\curl\lb \bfv\ts \bfb\rb$.   {\it Thus, 
some way of distinguishing the small-scale
quantities in the turbulent EMF from those in terms of which 
it is expanded using (\ref{BPRIME1a}) is necessary}. 

We will first formulate the required modifications
to the standard theory.  Then we study their effects in attempting to derive
the  Reynolds number dependence of  the back reaction on  $\alpha$ in the
limit of weak mean field.
%Growth of $\bbb$ 
%$\bbw$ 
%requires the $\nabla\times$ terms in
%(\ref{BBAR}) and (\ref{OMEGABAR}) to be non-vanishing.  However,
%when the turbulence is strictly  homogeneous in $\bfvp$ and $\bfbp$,
%these quantities vanish.  For purely isotropic turbulence, the term
%$\lb \bfvp\times\bfbp\rb$ in  (\ref{BBAR}) vanishes since it is
%the average of a vector, while $\curl \lb\bfvp\ts\bfwp\rb\propto
%\curl \nabla\lb v'^2\rb$ and  $\curl \lb\bfbp\cdot\nabla
%\bfbp\rb\propto \curl \nabla \lb b'^2\rb$ vanish from Reynolds rules
%and incompressibility.  Thus, anisotropy and inhomogeneity must be
%present for nontrivial time evolution of mean fields.
To this end, we split up the equation for the small-scale 
magnetic field into an equation for
that component which is  independent of the mean magnetic field
and that component which depends on the mean magnetic field. 
% \cite{BLA96} considers an expansion in $\bbv$ as well computing
%vorticity as well as mean field growth.  
Blackman \& Chou (1997) applied
this approach to first order in both the mean field and the velocity
field to find a coupled vorticity-magnetic dynamo. With a similar approach,
Field et al. (1998)  compute rigorously the effect of the back reaction of
$\bar{\bf B}$ on the dynamo
$\a$ to all orders in $\bar B$; here we consider only the first-order
terms in their development.  

We assume that the turbulence is weakly anisotropic and 
inhomogeneous.  Terms linear in the slowly time-varying mean quantities
contribute, but their averaged zeroth-order coefficients  are taken to be
isotropic and homogeneous (still allowing for reflection asymmetry).
Iterating the equations using the formal solutions for the turbulent fields
$\bfb(t)=\bfb(t=0)+\int \partial_{t'}\bfb dt'$ and
$\bfv(t)=\bfv(t=0)+\int \partial_{t'}\bfv dt'$, and using times
appropriately chosen such that the correlation
%$\lb\bfv(t)\ts\bfwp(0)\rb= 
$\lb\bfv(t)\ts\bfb(0)\rb \simeq 0$, we 
obtain to first order in mean quantities the turbulent EMF
\begin{equation}
\lb\bfv\ts\bfb\rb^{(1)}
=\lb\bfv^{(0)}(t)\ts\int^t_0\partial_{t'} 
\bfb^{(1)} dt' \rb-\lb\bfb^{(0)}(t)\ts\int_0^t \partial_{t'} 
\bfv^{(1)} dt' \rb,
\label{VPWP}
\end{equation}
where the time derivatives are given by (3) and by the equation of motion,
and thus to first order, (\ref {BBAR}) becomes 
\begin{equation}
\partial_{t}\bbB =\curl\lb\bfv\ts\bfb\rb^{(1)}
+\la\nabla^{2}\bar{\bf B}.
\label{BBAR2}
\end{equation}
Here  another 
 ambiguity   arises if one does not properly
separate the zeroth-order quantities from the higher order quantities: 
Note that the 
quantity $\lb\bfv\ts\bfb\rb$ can be written as
$\lb\bfv\ts\bfb\rb=\lb\bfv(t)\ts \int\partial_{t'}\bfb(t')dt'\rb=
-\lb\bfb(t)\ts \int \partial_{t'}\bfv(t')dt'\rb$.
But not
distinguishing between $\bfv,\bfb$ and $\bfv^{(0)}, \bfb^{(0)}$ in
(\ref{VPWP})   leads to
$\lb\bfv\ts\bfb\rb=
\lb\bfv(t)\ts\int \partial_t\bfb(t')dt'\rb=
-\lb\bfb(t)\ts \int \partial_t\bfv(t')dt'\rb=
\lb\bfv(t)\ts \int\partial_t\bfb(t')dt'\rb-\lb\bfb(t)\ts \int 
\partial_t\bfv(t')\rb=0$,
where the second last equality follows from purposely ignoring
the superscript $^{(0)} $ in (6), and the last equality is therefore 
unavoidable.  This shows that separating the zeroth order fluctuating
quantities from the total small-scale quantities is essential.

According to (6) the calculation of the EMF
requires expressions for $\bfb^{(1)}$ and
$\bfv^{(1)}$.  
The equation that determines $\bfv^{(1)}$ is the momentum
equation, given in general by 
\begin{equation}
\partial_{t} {\bfV}=-\bfV\cdot\nabla\bfV
-\nabla p_{\em{eff}}
 + \nu\nabla^{2}{\bf V}+{\bf B}\cdot\nabla{\bf B}+{\bf G}
+{\bf F}({\bf x}, t),
\label{NS}
\end{equation}
\noindent Here 
${\bf F}$ is the force that drives the turbulence, 
$p_{\em{eff}} \equiv p+{1\over2}B^2$, with $p\equiv P/\rho$, 
$\nu$ is a constant viscosity,
and $\bf G$ is gravity.
%The $\nabla\phi$ includes potential forces such as gravity.
To put this equation in the form that we need, we first define the small-scale
velocity $\bfv$ to be $\bfV-\bbV$ and similarly for the other variables. Then
averaging the resulting equation for $\bbV+\bfv$ and subtracting the result 
from (\ref{NS}), we obtain
\begin{equation}
\begin{array}{r}
\partial_{t} \bfv =
-\bbV\cdot\nb\bfv
-\bfv\cdot\nb\bbV
-\bfv\cdot\nb\bfv+
\lb\bfv\cdot\nabla\bfv\rb\\
%-\bfvp\cdot\nabla\bfvp
%-\bbv\cdot\nabla\bfvp
%-\bfvp\cdot\nabla\bbv_T
-\nabla p
-\nabla{1\over2}b^2
+\nabla \lb{1\over2} b^2\rb 
-\nabla( \bfbp\cdot \bbB)\\
+\bfbp\cdot\nabla\bfbp
-\lb\bfbp\cdot\nabla\bfbp\rb
+\bfbp\cdot\nabla\bbB
+\bbB\cdot\nabla\bfbp 
+{\bf f}({\bf x}, t)
+\nu\nabla^{2}\bfvp,
\label{VPRIME}
\end{array}
\end{equation}
where $p$ is now the small-scale pressure and $\bf f$ is the small-scale
(and only) part of $\bf F$; consistent with the assumption of 
incompressibility, we take the small-scale gravity to be zero. 
We eliminate the term $\bbV\cd\nb\bfv$ by a change of frame. 
We shall also ignore the term $\bfv\cd\nb\bbV$.  In our Galaxy for example, 
this term is smaller than $\vert\bfv\cd\nb\bfv\vert$ (Field et al 1998).
In what follows, therefore, $\bbV$ will not appear. 
Now, writing the equations (\ref{BPRIME}) and  (\ref{VPRIME}) 
to zeroth and linear order in  
$\bbB$, we have 
\begin{equation}
\begin{array}{r}
\partial_{t} \bfbp^{(0)} = \bfbp^{(0)}\cdot\nabla\bfvp^{(0)} 
-\bfvp^{(0)}\cdot\nabla\bfbp^{(0)}
-\curl\lb\bfvp^{(0)}\ts\bfbp^{(0)}\rb 
+ \la\nabla^{2}\bfbp^{(0)}
\end{array}
\label{BPRIME2}
\end{equation}
and
\begin{equation}
\begin{array}{r}
\partial_{t} \bfbp^{(1)} =
\bbB \cdot\nb\bfvp^{(0)} -\bfvp^{(0)} \cdot\nb\bbB+
\bfbp^{(1)} \cdot\nabla\bfvp^{(0)}  -\bfvp^{(1)} \cdot\nabla\bfbp^{(0)} 
\\
+ \bfbp^{(0)} \cdot\nabla\bfvp^{(1)}  -\bfvp^{(0)} \cdot\nabla\bfbp^{(1)} 
-\curl\lb\bfvp\ts\bfbp\rb^{(1)}  
+\la\nabla^{2}\bfbp^{(1)} .
\end{array}
\label{BPRIME3}
\end{equation}
for the small-scale magnetic field, and 
\begin{equation}
\begin{array}{r}
\partial_{t} \bfvp^{(0)}=
-\bfvp^{(0)}\cdot\nabla\bfvp^{(0)}
+\lb\bfvp^{(0)}\cdot\nabla\bfvp^{(0)}\rb
%-\bbv\cdot\nabla\bfvp
-\nabla p^{(0)}
-\nabla {1\over2}{\b0}^2
+\bfbp^{(0)}\cdot\nabla\bfbp^{(0)}
-\lb\bfbp^{(0)}\cdot\nabla\bfbp^{(0)}\rb\\
+\nabla \lb {1\over2}{\b0}^2\rb 
+{\bf f}({\bf x}, t)^{(0)}
+\nu\nabla^{2}\bfvp^{(0)},
\label{VPRIME0}
\end{array}
\end{equation}
and
\begin{equation}
\begin{array}{r}
\partial_{t} \bfvp^{(1)} = 
-\bfv^{(0)}\cd\nb\bfv^{(1)}
-\bfv^{(1)}\cd\nb\bfv^{(0)}
-\nb p^{(1)}
-\nb(\bfb^{(0)}\cd\bfb^{(1)})
+\nb\lb\bfb^{(0)}\cd\bfb^{(1)}\rb
+\lb\bfvp\cdot\nabla\bfvp\rb^{(1)}
%-\bfvp\cdot\nabla\bfvp
%-\bbv\cdot\nabla\bfvp
-\nabla( \bfbp^{(0)}\cdot \bbB)\\
+\bfbp^{(1)}\cdot\nabla\bfbp^{(0)}
+\bfbp^{(0)}\cdot\nabla\bfbp^{(1)}
-\lb\bfbp\cdot\nabla\bfbp\rb^{(1)}
+\bfbp^{(0)}\cdot\nabla\bbB
+\bbB\cdot\nabla\bfbp^{(0)} 
+\nu\nabla^{2}\bfvp^{(1)},
\label{VPRIME1}
\end{array} 
\end{equation}
for $\bfvp^{(1)}$, where we have assumed that the forcing function is the same for 
both the state with $\bbB=0$ and that with $\bbB\not=0$, so ${\bf f}^{(1)}=0$.

To further simplify we need an approximation related to, but
distinct from the standard FOSA.
Note that the standard FOSA, employed in section 1, 
would imply that 
$\vert\bfbp\vert =\vert\bfbp^{(0)}+\bfbp^{(1)}+...\vert< 
\vert{\bar {\bf B}}\vert$.
Although this assumption is different from the kinematic approximation in that
it says nothing about the back reaction of the mean field
on the velocity flows, it does imply that the
small scale field is much weaker than the mean field. In reality, 
this is not the case:  both simulations and analytical arguments
(e.g. Parker 1979) show that the small-scale field energy saturates
at a value of order the small-scale kinetic energy .  Fortunately, 
we do not have to make the FOSA here.  This is because we 
take the zeroth-order turbulent quantities to be solutions of (10) and (12), 
and work with the effect of the mean field on the first order quantities.
%(all orders are considered in Field et al. 1998). 
We require only that $\vert\bfbp^{(1)}\vert,\vert\bfvp^{(1)}\vert<\bB$, 
a much weaker condition.
This reduced smoothing approximation (RSA), unlike the FOSA,
allows for $b^{(0)}>>\bB$.
% which is known to pervade
%when the mean field is weak because the total small
%scale field (dominated by $\bfbp^{(0)}$) builds up to equipartition
%on a time scale $\sim t_c^2/t_{ed}$ where $t_c$ and $t_{ed}$ are the
%correlation and eddy turnover times.  

This RSA allows us to eliminate terms second order 
in the fluctuating quantities in (\ref{VPRIME1}) and (\ref{BPRIME3}).  
Note, however, 
that in the first $\sim t_c$ of growth, the first term in (\ref{VPRIME1})
could dominate the terms with the mean field.
However, we are only interested in times long compared to this time, after
which a steady state ensues.
When we put the equation
for $\bfvp^{(1)}$ into the average (\ref{VPWP})
none of the bracketed terms
on the right of (11) or $(\ref{VPRIME1})$ contribute. 
The relevant equations for the small-scale fields then become
\begin{equation}
\begin{array}{r}
\partial_{t} \bfvp^{(1)} = 
%-\bfvp\cdot\nabla\bfvp
%-\bbv\cdot\nabla\bfvp
-\nabla p^{(1)}
-\nabla( \bfbp^{(0)}\cdot \bbB)
+\bfbp^{(0)}\cdot\nabla\bbB
+\bbB\cdot\nabla\bfbp^{(0)} 
+\nu\nabla^{2}\bfvp^{(1)},
\label{VPRIME2}
\end{array}
\end{equation}
and 
\begin{equation}
\begin{array}{r}
\partial_{t} \bfbp^{(1)} =
\bbB \cdot\nb\bfvp^{(0)} 
-\bfvp^{(0)} \cdot\nb\bbB+ 
\la\nabla^{2}\bfbp^{(1)} .
\end{array}
\label{BPRIME4}
\end{equation}

A remaining problem is to deal with the 
pressure in (\ref{VPRIME2}).  
%following Blackman \& Chou (1997)
For present purposes, 
we focus on the correction
to the $\alpha$ dynamo coefficient, and so 
we consider the special case where $\nabla\bB$ is negligible
(Field et al. 1998). We can then take the divergence of (14). The
resulting equation is $\nabla^2 (p^{(1)}+ \bfbp^{(0)}\cdot \bbB)=0$,
whose solution is $(p^{(1)}+ \bfbp^{(0)}\cdot \bbB)=constant$.
Thus the first two terms drop out of (14).
%Thus we can follow Field et al (1998) and simply deal with the
%pressure.  
%(Blackman \& Chou (1997) use a homogeneity and isotropy 
%argument to suggest that the pressure does not contribute 
%to first order in the mean fields even when mean field 
%gradients cannot be ignored.)

Approximating time integrals in 
%$\lb\bfwp\ts\bfbp\rb^{(1)},\, \lb\bfbp\cdot\nabla
%\bfbp\rb^{(1)}$ and 
$\lb\bfvp\times\bfbp\rb^{(1)}$ 
by factors of the correlation time $t_{c}$ 
%(\cite{RUZ88}) 
and freely
employing Reynolds rules and incompressibility, we then obtain
\begin{equation}
\begin{array}{ll}
\lb\bfvp\ts \bfbp\rb^{(1)}=\a^{(0)}\bbB 
\label{MESS1}
\end{array}
\end{equation}
where 
\begin{equation}
\begin{array}{ll}
\a^{(0)}
=-{1\over3}t_c[\lb\bfvp^{(0)}\cdot\curl\bfvp^{(0)}\rb
-\lb\bfbp^{(0)}\cdot\curl\bfbp^{(0)}\rb].
\label{ALPHA}
\end{array}
\end{equation}

%\begin{equation}
%\begin{array}{ll}
%\beta^{(0)}_{ijk}=\beta^{(0)}\ep_{ijk}=
%(t_c/3)[\lb\bfvp^{(0)}\cdot\bfvp^{(0)}\rb+
%2\lb\bfbp^{(0)}\cdot\bfbp^{(0)}\rb]\ep_{ijk}.
%\label{BETA}
%\end{array}
%\end{equation}
%\displaystyle
%(3/\tau_{c})\curl\lb\bfvp\ts\bfbp\rb^{(1)}
%= &\displaystyle \lb \v0_{i}\w0_{i}\rb(\curl\bbw) +
%\lb\v0_{i}\v0_{i}\rb
%\nabla^{2}{\bbw} + 
%2\lb\w0_{i}\b0_{i}\rb\nabla^{2}\bbb  + \\
%\: & \displaystyle \quad 
%\lb\epsilon_{ijk}\w0_{k}\partial_{i}\b0_{j}\rb(\curl\bbb)
%-\lb\v0_{i}\b0_{i}\rb\nabla^{2}(\curl\bbb)
%\noindent with similar expressions for
%$\curl\lb\bfbp\cdot\nabla\bfbp\rb^{(1)}$ and 
Upon substituting these into (\ref{BBAR}), the curls can be pulled onto
the  $\bbB$ from homogeneity of the zeroth-order
averages.  
%After some simplification, we get 
%\begin{equation}
%d_{t}\bbw = \alpha_{0}(\curl \bbw) +
%\alpha_{1}(\curl\bbb) +\beta_{0}\nabla^{2}\bbw
%+\beta_{1}\nabla^{2}\bbb - \alpha_{2}\nabla^{2}(\curl\bbb)
%+\bbw_c\cdot\nabla\bbv
%+\bbw\cdot\nabla\bbv_c+\bbw_c\cdot\nabla\bbv_c,
%\label{WEQM}
%\end{equation}
%\begin{equation}
%\partial_{t}\bbb =
%\alpha^{(0)}(\curl\bbb)
%+\beta^{(0)}\nabla^{2}\bbb
%\bbb\cdot\nabla\bbv_c
%\label{BEQM}
%\end{equation}
Note that we have ignored the sub-dominant $\nu$ and $\la$
terms. This will not pose any problem for our discussion 
in the next section: the entrance of the Reynolds number 
into the theory that we will address is that purported 
to enter through  the ``magnetic correction''
$\lb\bfbp^{(0)}\cdot\curl\bfbp^{(0)}\rb$ in (\ref{ALPHA}).
Similar magnetic corrections to that in (\ref{ALPHA})
have been derived before 
(e.g. Pouquet et al., 1976; Gruzinov \& Diamond 1994) but none
of the previous papers distinguish between $\a$ and $\a^{(0)}$,
and we  will see below that the difference is fundamental.

%Note also that the factors of 2 in (\ref{ALPHA}) and (\ref{BETA})
%relate to our treatment of the pressure.

\centerline{\bf 4. Contrast to previous work}

Building on the ideas of e.g. Piddington (1981), 
most workers agree that the Lorentz forces 
from the growing magnetic field react back on the turbulent 
motions driving the field growth and  
complicating the turbulent motions.
However, analytic studies and simulations disagree as to the extent to
which the dynamo coefficients are suppressed by this back reaction.
Some (e.g. \cite{CAT91}, \cite{VAI92},
\cite{CAT94}, \cite{CAT96}, \cite{GRU94}, \cite{BHA95}, \cite{KLE95}) 
argue that the suppression of  
e.g. $\a$ takes the form 
$\a\sim \a^{(0)}/(1+R_M\bB^2/v^2)$
where $R_M$ is the magnetic Reynolds number, 
while others (e.g. Kraichnan 1979, Field et al., 1998) suggest 
$\a\sim \a^{(0)}/(1+\bB^2/v^2)$ in the fully dynamic regime.
If the former formula represented the actual level of suppression, the large
values of $R_M$ in nature would prevent astrophysical dynamos from working.  
Something else would be needed to generate large-scale fields in stars and
galaxies.  More analytic and numerical studies will ultimately be required to 
answer this question.
At present, we suggest 
that some existing analytic arguments for extreme 
suppression can be challenged. 

%\centerline {\bf 4.1 On arguments suggesting strong suppression of 
%helicity}

To do so, we follow the path of \cite{GRU94} and \cite{BHA95} but
employ the formalism of section 3, rather than that of section 2, 
and then find that we do not arrive at the same conclusions.  
Consider the small-scale electric field:
\begin{equation}
\bfe= c^{-1}[\lb \bfv \ts \bfb\rb
-\bfv\ts\bfb
-\bfv\ts\bbB]
+\eta \bfj,
\label{BFE1}
\end{equation}
where $\eta = 4\pi\lambda/c^2$ is the resistivity.
Dotting this with $\bfb$ and averaging gives
\begin{equation}
\lb\bfb\cdot\bfe\rb=-c^{-1}\lb\bfb\cdot (\bfv\ts \bbB)\rb
+\eta \lb \bfb\cdot\bfj\rb,
\label{BFE2}
\end{equation}
where we note that the first two terms on the 
right of (\ref{BFE1}) do not contribute to (\ref{BFE2}).
Using the triple product on the second term in (\ref{BFE2}) and rearranging
gives:
\begin{equation}
\lb\bfb\cdot\bfj\rb=
\eta^{-1}[-\bbB\cdot\lb\bfv \ts \bfb/c\rb
+\lb\bfb\cdot\bfe\rb].
\label{BFE3}
\end{equation}
Now the ``magnetic correction'' to $\a$ 
in the linear regime as computed in (17)
is proportional to  $c\lb\bfb\cd\curl\bfb\rb^{(0)}/4\pi=\lb\bfb\cdot \bfj\rb^{(0)}$,
not to $\lb\bfb\cdot \bfj\rb$.  It is very important to distinguish
between those quantities with and without the superscript $(0)$, as we now 
show.
Taking the zeroth order contribution to (\ref{BFE3}) gives 
\begin{equation}
\lb\bfb\cdot \bfj\rb^{(0)}=\eta^{-1}\lb\bfb\cdot \bfe\rb^{(0)},
\label{BFE4}
\end{equation}
but using (\ref{BFE1}) for $\bfe^{(0)}$ in (\ref{BFE4}) 
shows that (\ref{BFE4}) is a trivial identity.
Thus it is not possible to get information from (\ref{BFE3}).
If we do not distinguish the zeroth-order quantities from first-order
quantities we might conclude that the
left hand side of (\ref{BFE4}) is equal to the correction to 
$\a$ derived in (\ref{ALPHA}), but that is not correct.

An almost identical discussion applies to Kleeorin et al (1995). In their 
Appendix A they provide a somewhat similar derivation of the
back reaction to $\a$, with the resulting equation 
\beq
\partial_t\a_m=C_1(\bbB\cdot\curl\bbB-\a\bbB^2/\beta)-C_2\a_m,
\label{KLE1}
\ee
where $C_1$ and $C_2$ are defined constants,
and they write $\a=\a_0+\a_m$,  with their 
$\a_0\propto \lb\bfvp\cdot\curl\bfvp\rb$ and
$\a_m\propto \lb\bfbp\cdot\curl\bfbp\rb$.
The subtlety arises from their statement
below their un-numbered equation between their (A4) and (A5) 
that ``$\a$ and $\eta_T(\equiv\beta)$ are scalars when the mean field is small.''
Scalars result from the assumption of isotropy, but isotropy applies 
only to strictly zeroth-order quantities as
discussed in section 3.  Thus, the ambiguity in the notation 
discussed above is present.  We suggest that 
the $\a$ and $\eta_T(\equiv \beta)$ to which Kleeorin et al (1995) refer
should  actually be 
$\a^{(0)}$ and $\eta_T^{(0)}(\equiv\beta^{(0)})$.
Thus, the magnetic correction to $\a$ in the weak mean field limit
which they define as $\a_m$, should really be 
$\a_m^{(0)}\propto\lb\bfbp^{(0}\cdot\curl\bfbp^{(0)}\rb$, 
not $\propto \lb\bfbp\cdot\curl\bfbp\rb$.
This latter quantity is not simply related to $\a_m^{(0)}$.
The distinction is essential.  Rather than (\ref{KLE1}), 
the equation for the correction would instead be 
\beq
\partial_t\a_m^{(0)}=-C_2\a_m^{(0)},
\label{KLE2}
\ee
where no mean field appears.  
As a result, (\ref{KLE1})  
implicitly contains less information than it
appears to, by analogy to (\ref{BFE3}-\ref{BFE4}).
\medskip

Yet another subtlety arises when one notices that the right
side of (\ref{BFE4}) is proportional to the sum of a time 
derivative and spatial divergence of averages
of zeroth order correlations.
Assuming exact stationarity and isotropy of zeroth order
averages then implies $\lb\bfb\cdot \bfe\rb^{(0)}=0$.
The left side of (\ref{BFE4}) would then also be zero, 
and thus the correction in (\ref{BFE4}) would vanish
for any finite $\eta$. 
The current helicity would then not be a correction to $\a$ but
would be related directly through (\ref{BFE3}).
However, the fact that the diffusivity enters (\ref{BFE4}) 
means that the right side is the ratio of two small quantities
and requires an assessment of how accurate the assumptions
of stationarity and isotropy really are.

The vanishing of $\lb\bfb\cdot \bfe\rb^{(0)}$
is reminiscent of Seehafer (1994) and Keinigs (1983), but 
is different because these authors do not distinguish 
zeroth order quantities and Keinigs (1983) employs FOSA.
Since averages of non-zeroth 
order fluctuating quantities are not in general isotropic, homogeneous
or stationary, the full 
$\lb\bfb\cdot \bfe\rb\ne 0$.  When the full dynamo equation
is considered, mean quantities that are functions of non-zeroth
order fluctuating quantities can vary on the large scale of 
the system. The second term on the right of (\ref{BFE3})
can exceed the left hand side, providing the dominant
balance to the first term on the right.
This is not apparent from Keinigs (1983) or Seehafer (1994) because they
do not distinguish between zeroth-order and higher order
small-scale quantities.  

\centerline{\bf 5. Discussion and Conclusions}

We have pointed out a fundamental ambiguity
in magnetic dynamo theory, resolved it, and have shown 
its importance in explaining why some analytic 
approaches to the problem of the back reaction of the mean field 
in dynamo theories disagree.  
In particular, we showed that only part of the 
small-scale quantities can be thought of as isotropic, no matter how small 
the mean field is, and why the formalism of 
standard treatments is ambiguous on this point.
%Taking the helicity dynamo coefficient as an example,
The ``magnetic correction'' term shown in (\ref{ALPHA}) 
is constructed from only that component of the small-scale field which is 
independent of the mean field, not from the total small-scale field.
Previous studies have not separated the small-scale quantities
into their zeroth and higher order components, and as a
a result find that the correction term in (\ref{ALPHA})  
appears to relate to the magnetic Reynolds number through (\ref{KLE1})
or (\ref{BFE3}).  When the separation is made, 
the correction term cannot be interpreted as a strong suppression to the 
$\a$ effect because then the only place 
for the magnetic Reynolds number to enter is through the last term
in (\ref{BPRIME4}). This does not produce a strong 
suppression of $\a$ (Field et al 1998).  Note that the first-order smoothing
approximation is not necessary in our approach.

The fact that a true dynamo cannot be kinematic
need not imply that dynamo action is suppressed.
An often quoted ratio, that of the total magnetic energy to that contained
in the mean field (e.g. \cite{ZEL83}) $\lb{\bf b}^2\rb=R_M^n\lb\bB^2\rb$,
where $n$ is some non-zero power 
(or logarithmic for a purely Kolmogorov spectrum, applicable in the
completely kinematic regime), 
applies only when the first two terms on the right
of (\ref{BPRIME}) approximately balance the next three, or equivalently,
when the sum of the the first three terms on the right
of (\ref{BPRIME2}) balance the first two terms on the right of 
(\ref{BPRIME3}), since $\partial_t\bfbp=\partial_t\bfbp^{(0)}+
\partial_t\bfbp^{(1)}$.  But this state is very short lived because
the small-scale field rapidly grows to near equipartition with the turbulent
energy on a time scale $\sim t_c$, and the first three terms
on the right of (\ref{BPRIME2}) quickly dominate.  These terms
lead to an MHD turbulent spectrum with most of the energy
contained on the outer scale of the turbulence.  In principle,
the mean field can still slowly grow, while the combination
of field line stretching and nonlinear damping in a turbulent 
cascade maintain the small-scale field at a steady energy density.
A key question (\cite{FIE95}) is: what is the integrated
kinetic vs. magnetic energy on the outer scale of the turbulence?
We believe that this ratio is essential in assessing whether the dynamo 
works, and suggest that a dynamo 
can work as long as there is a mismatch. 
%in either direction,
%corresponding to dominance of either of the two terms in (\ref{ALPHA}).

Finally, note that 
the growth of the small-scale field necessarily means that even
galaxies at redshifts $\gsim 2$ would exhibit substantial fields 
when observed by Faraday rotation.
This is because a typical turbulent energy-containing eddy in a galaxy
is $\sim 100$pc across, so that any line of sight through a disk will have
significant RMS mean field component (Blackman 1998).  A rough lower limit 
comes from assuming that the 
small-scale field builds up to equipartition with the turbulent
motions ($\sim {\rm few} \ts 10^{-6}$G in our Galaxy).
Even when observed edge on, there are only $\sim 100$ such cells along the 
line of sight. The observed mean field would then be $\sim 10$ times smaller, 
or $\gsim 10^{-7}$G.  
The mere existence of apparent mean fields is therefore not a strong
test for dynamo theory even for large redshift Galaxies. 
A better observational test would be to measure the scale and
pattern of field reversals at these redshifts.  The
existence of small-scale fields is guaranteed to accompany any
large-scale field produced by dynamo action and does not preclude 
its operation.

\vfill
\eject

Acknowledgements: G.F. acknowledges partial support from NASA Grant NAGW-931.

\end{document}